\documentclass{amsproc}
\usepackage[english]{babel}
\usepackage[useregional]{datetime2}  
\usepackage{graphicx}
\usepackage{hyperref}
\usepackage{appendix}

\usepackage{amscd}
\usepackage{amsmath}
\usepackage{epsfig}
\usepackage{amsfonts}
\usepackage{amssymb}
\usepackage{hyperref}
\usepackage{cite}

\usepackage{pdfpages}
\usepackage[thinc]{esdiff}
\usepackage{accents}
\usepackage{nicefrac}

\theoremstyle{definition}

\theoremstyle{remark}

\numberwithin{equation}{section}

\definecolor{auburn}{rgb}{0.43, 0.21, 0.1}
\definecolor{burgundy}{rgb}{0.5, 0.0, 0.13}
\definecolor{magenta}{rgb}{1.0, 0.0, 1.0}
\definecolor{magenta(dye)}{rgb}{0.79, 0.08, 0.48}
\definecolor{balmerred}{RGB}{220,20,60}      
\definecolor{balmerbluegreen}{RGB}{0,170,170} 
\definecolor{balmerblue}{RGB}{65,105,225}     
\definecolor{balmerviolet}{RGB}{138,43,226}   

\usepackage[svgnames]{xcolor}

\allowdisplaybreaks
\input{tcilatex}

\newcommand{\iun}{\mathrm{i}\mkern1mu}

\begin{document}

\title[Remarkable Dates and a Place]
{Schr\"{o}dinger's Wave Mechanics: Remarkable Dates \\ and Place One Hundred Years Ago\/}
\author{Sergei K. Suslov}
\address{School of Mathematical and Statistical Sciences, Arizona State
University, P.~O.\ Box 871804, Tempe, AZ 85287-1804, U.S.A.}
\email{sergei@asu.edu}
\date{August 24, 2026}

\begin{abstract}
We discuss, at a level accessible to a wide audience, including students and teachers of physics and mathematics, 
the fundamental transition from classical mechanics to wave equations made by Schr\"{o}dinger a century ago.
These historical events clarify the structure and significance of quantum mechanics 
and are of interest to the global scientific community.
Valuable sources for further studies are provided throughout the article. 
\end{abstract}

\dedicatory{\begin{center}
{\scriptsize{Dedicated to the 100th anniversary of the birth of wave mechanics}}
\end{center}}

\maketitle

\begin{center}
{{\scriptsize{Compiled on:  December 29, 2025 at 12:05~am Arosa Switzerland time  \\ 
Amended on: December 30, 2025 at 8:29~am Munich Germany time \\
Last amended on: August 25, 2026 at 6:02~am Arizona USA time \\
}} }%
\end{center}

\bigskip
\bigskip
\noindent
Exactly a century ago, wave quantum mechanics was born in Arosa, Switzerland.
Erwin Schr\"{o}dinger was vacationing in this classic Swiss Alps town at Christmas 1925 
when he made his breakthrough discovery of the wave equation \cite{SchrQMI}.

\bigskip
\noindent
How did Schr\"{o}dinger derive his celebrated wave equation a century ago and subsequently apply it to the hydrogen atom? According to his own testimony  \cite{SchrQMI, SchrodingerCohrent, SchrPhysRev1926} and \cite[030\dag\;~pp.~141--143]{Meyenn2011},
de Broglie’s seminal work on the wave theory of matter (1923–24) and Einstein’s studies on ideal Bose gases (1924–25) 
laid the foundation for the discovery of wave mechanics%
\footnote
{Schr\"{o}dinger is more specific in \cite{SchrodingerCohrent} and in his letter to Lorentz on March 30, 1926 \cite[055\dag\;~pp.~203--205]{Meyenn2011}: 
``... {\textit{I would like to add the following remarks. The suggested approach originates from the insightful theses of L.~de~Broglie, {{\rm{Annales de Physique}} (10) {\bf{3}}, 22, 1925} $(${\rm\cite{deBroglie1925}, {SKS}}$\ )$
and the interesting remarks of A.~Einstein, {\rm{Berliner Berichte},} S.~9ff., 1925 $(${\rm{\cite{Einstein24-25a, Einstein24-25b}, SKS}}$\ )$}}...''$\ .$
(See also his remarks on p.~9 in the first `quantization' article \cite{SchrodingerBook}.)
}
(see also \cite{MEHRAIII, Rechenberg} and \cite{Barleyetal2021}).

\bigskip
\noindent
In his letter to Einstein dated November 3, 1925, Schr\"{o}dinger writes:
``{\textit{A few days ago, I read with great interest the ingenious theses of Louis de Broglie, which I finally got hold of ...\;}"
} \cite[030\dag\;~p.~142]{Meyenn2011} .
A few months later, in his letter to Einstein on April 23, 1926 \cite[062\dag\;~p.~215]{Meyenn2011}, Schr\"{o}dinger admits: ``  {\textit{Incidentally, the whole thing would certainly not have come about now, and perhaps never would have $(I$ mean, not from my side$)$, if your second work on gas degeneration hadn't made me aware of the importance of de Broglie's ideas.''}}

\vfill
\eject
\newpage
\medskip
\noindent
{\scshape{Louis de Broglie testimony:\/}}
{{``{\it{After long reflection in solitude and meditation, I suddenly had the idea, during the year 1923, that the discovery made by Einstein in 1905 should be generalised by extending it to all material particles and notably to electrons}}''\cite{deBroglie1967}}.}
%
%

\medskip
\noindent
{{The connection with Einstein is as follows:\/} 
de Broglie generalized Einstein’s 1905 idea that light behaves as particles (photons) by proposing the reverse -- that particles (like electrons) behave as waves \cite{Jammer1989}.}

\medskip
\noindent
{\scshape{Timetable:}}
The exact dates of Schr\"{o}dinger's foundational discoveries, leading to his first publications \cite{SchrQMI, SchrQMII}, are not precisely recorded \cite{Bloch1976, DebyeInterview, Kragh1982, Kragh1984, Wessels1979} and \cite[pp.~459--465]{MEHRAIII}.
However, one can estimate the timeline based on his letter to Einstein \cite[030\dag\;~pp.~141--143]{Meyenn2011}, dated November 3, 1925; Bloch’s recollection of two colloquia in Z\"{u}rich \cite{Bloch1976}, presumably held in late November and/or early December 1925 \cite[pp.~419--423]{MEHRAIII}; a letter to Wien \cite[037\dag~pp.~162--165]{Meyenn2011} from Arosa on December 27, 1925; and a letter to Sommerfeld \cite[041\dag~pp.~170--172]{Meyenn2011} from Z\"{u}rich on January 29, 1926.
This yields a reasonable estimate spanning from early November 1925 to the end of January 1926.

\medskip
\noindent
The discovery was the cornerstone of Schr\"{o}dinger's ``Annus Mirabilis'' (Miracle Year) and ultimately earned him the 1933 Nobel Prize in Physics.
Shortly after his Alpine retreat in Arosa,
in the pivotal letter to Arnold Sommerfeld from Z\"{u}rich Schr\"{o}dinger concludes: 
``{\textit{%
... Finally, I still wish to add that the discovery of the whole connection
$[$between the wave equation and the quantization of the hydrogen atom$]$ goes back to your beautiful quantization method for evaluating the radial quantum integral. It was the characteristic 
$- \frac{B}{\sqrt{A}} + \sqrt{C'}$, 
which suddenly shone out from the exponents ${\alpha_1}$ and ${\alpha_2}$ like a Holy Grail.%
}}''

\medskip
\noindent
This correspondence announced his revolutionary formulation of wave mechanics -- marking the birth of the Schr\"{o}dinger wave equation. 
In this letter, Schr\"{o}dinger reported for the first time the success of the new theory in solving the quantum oscillator, rotator, the non-relativistic (and partially relativistic) hydrogen atom (Kepler problems), and the free motion of a point mass in infinite space and in a box, prior to the formal publications \cite{SchrQMI, SchrQMII}.
He also formulated a program for future research. 
For the reader’s benefit, the complete letter has been translated from German to English in Appendix~D of our recent publication 
\cite{Bar:Ruf:Sus}.

\medskip
\noindent
In a letter dated February 3, 1926 \cite[042\dag~pp.~173--175]{Meyenn2011}, Sommerfeld responded enthusiastically: 
``{\textit{What you write, in your essay and letter, is terribly interesting. My personal opinion on the mysticism of integers must remain silent, as must my personal convenience ... My impression is this: Your method is a substitute for the new quantum mechanics of Heisenberg, Born, Dirac ...
Because your results are completely consistent with theirs...}}"

\medskip
\noindent
This marked the beginning of the triumph of Schr\"{o}dinger’s wave mechanics \cite[pp.~617--636]{MEHRAIII} (see also
\cite{EinsteinLetters, Fock1926, Jammer1989}).
\begin{figure}[hbt!]
\centering
\includegraphics[width=0.8\textwidth]{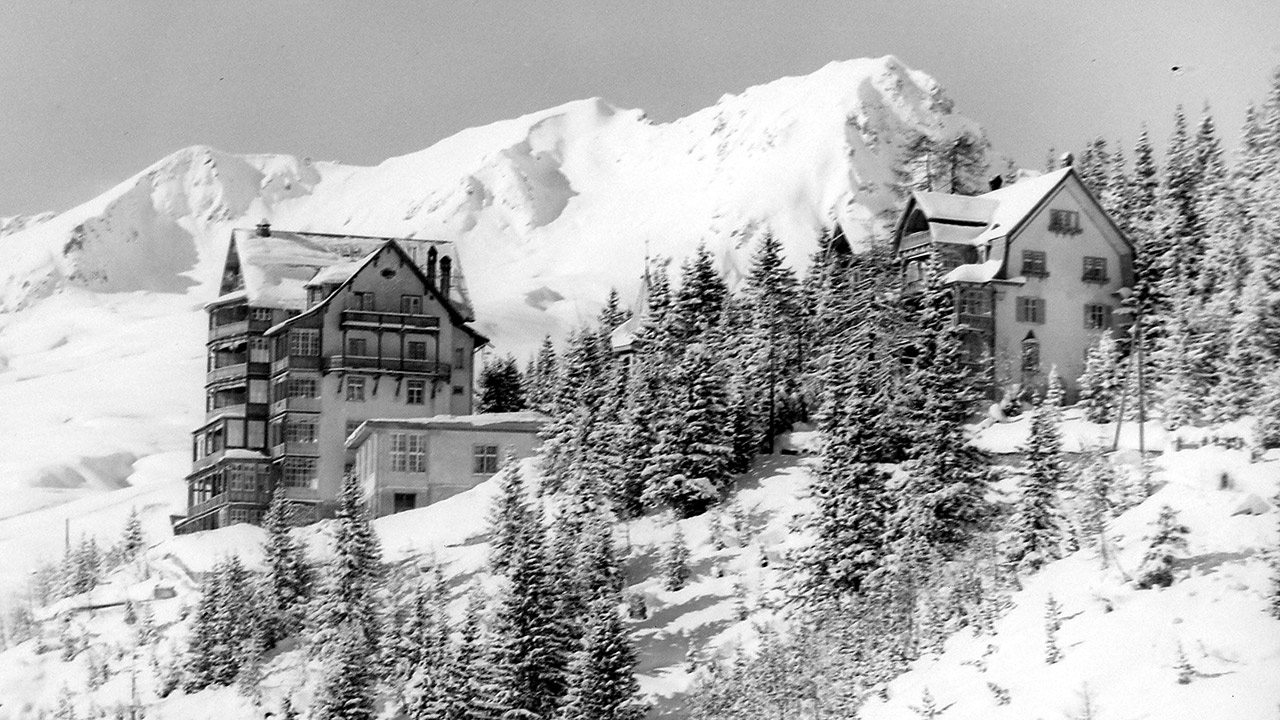}
\caption{The Villa Frisia of Dr.~Herwig's sanatorium, Arosa (right), where it is believed wave mechanics was discovered during the Christmas holidays 1925--26
\cite{Moore1989}.\/}
\label{Figure1}
\end{figure}

\begin{figure}[hbt!]
\centering
\includegraphics[width=0.8\textwidth]{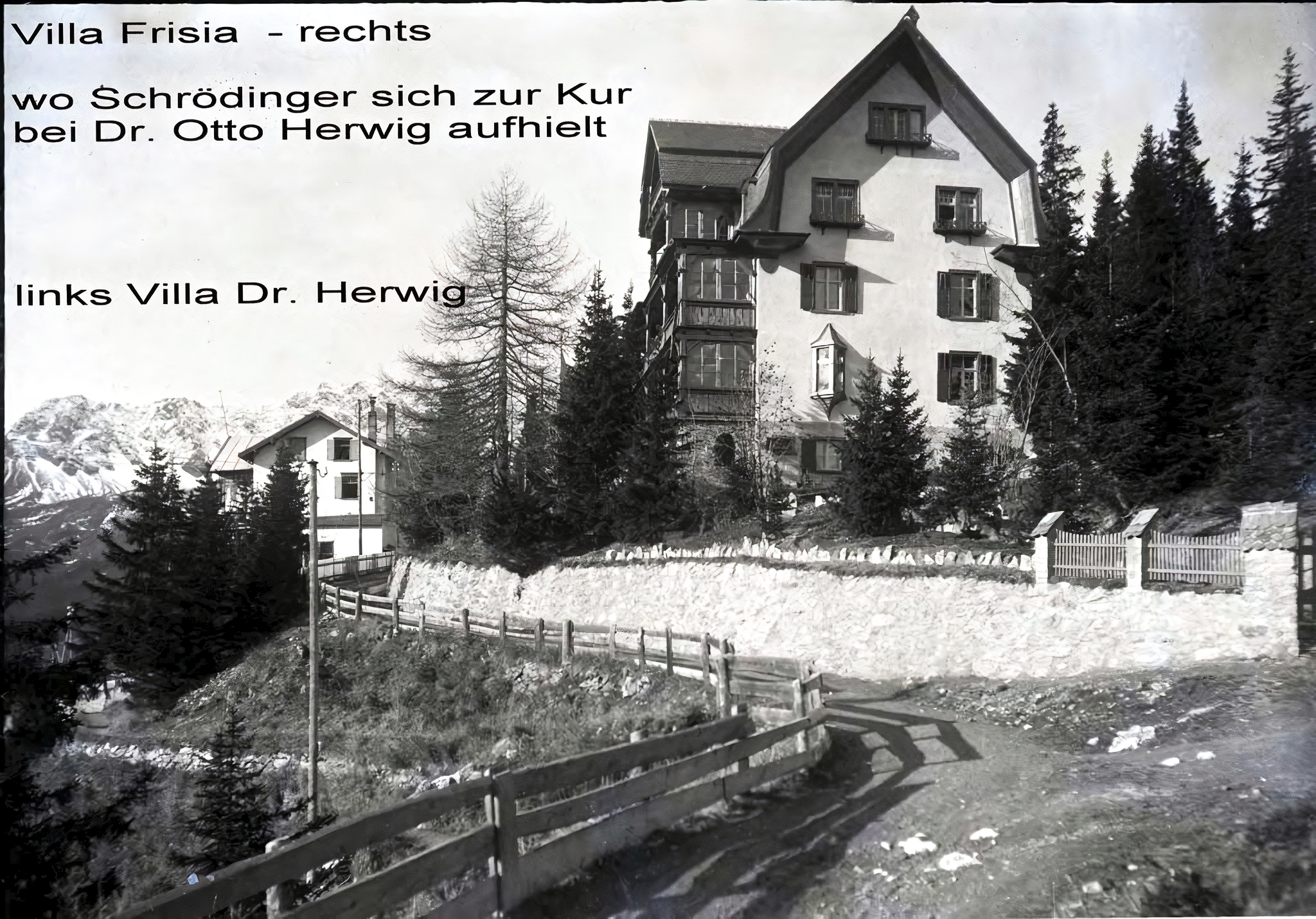}
\caption{A postcard with the view of the Villa Frisia (right), where Schr\"{o}dinger stayed, and the house of Dr.~Otto~Herwig (left).\/}
\label{Figure2}
\end{figure}
\medskip
\noindent
{\scshape{Place of discovery:}}
Arosa, an Alpine \emph{Kurort} at about 1800~m altitude, not far from ski-resort Davos, and overlooked by the great peak of the Weisshorn
(Figures 1 and 2).
For a related video, see: 
{\url{https://www.news.uzh.ch/en/articles/2017/Schroedinger.html} } 
Here, among other things, a female physicist, Professor Laura Badis, meets
the grandson of Dr.~Herwig and he shows her the
entry of the payment in a guest book, done by Schr\"{o}dinger\/.

\medskip
\noindent
{\scshape{Main legacy:}}
The time-dependent Schr\"{o}dinger equation, as shown in Figure~3, although
{\it{de facto}} required in an article dedicated to the coherent states \cite{SchrodingerCohrent}
\footnote{
The {\it{coherent states}}, or nonspreading wave packets, when the variances of coordinate and linear momentum are minimal \cite{Lan:Lif}, were introduced by Schr\"{o}dinger before inventing the time-dependent equation \cite{SchrodingerCohrent}. They also occurred in correspondence with Hendrik Lorentz \cite{EinsteinLetters, Moore1989}\/ -- originals in German: \cite[055\dag\;~pp.~203--205]{Meyenn2011}, \cite[073\dag\;~pp.~238--246]{Meyenn2011}, \cite[076\dag\;~pp.~252--261]{Meyenn2011} and with Max Planck
 \cite[074\dag\;~pp.~247--250]{Meyenn2011}
(see also \cite{Kryuchkov2013} for an extension to the minimum-uncertainty squeezed states when the variances oscillate under the Heisenberg limit).}
,
was published only about six months later \cite{SchrodingerTimeDependent}%
.

\begin{figure}[hbt!]
\centering
\includegraphics[width=0.85\textwidth]{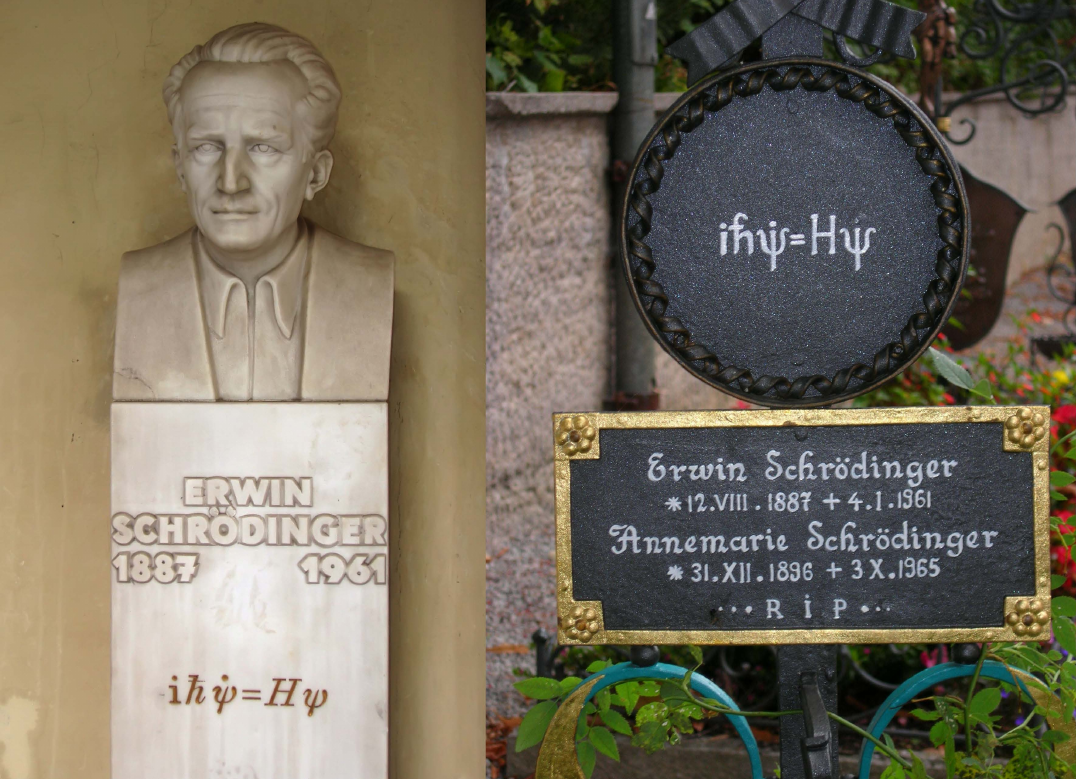}
\caption{The time-dependent Schr\"{o}dinger equation --- arcaded courtyard in the main building of the University of Vienna (left) and the churchyard cemetery of Alpbach village in Tirol (right), respectively. }
\label{Figure3}
\end{figure}
%
%
%

\medskip
\noindent
Schr\"odinger himself described the `second quantum revolution' as follows \cite{SchrodingerCohrent}:
``Building on ideas of {\scshape{de Broglie}} and {\scshape{Einstein\/}}, I tried to show
that the ordinary differential equations of mechanics, which attempt
to define the co-ordinates of a mechanical system as functions
of the time, are no longer applicable for ``small'' systems;
instead there must be introduced a certain {\textit{partial}} differential
equation, which defines a variable $\psi$ (``wave function'') as a function
of the co-ordinates and the time.''

\medskip
\noindent
The new fundamental law of nature, that describes the motion of electron in an atom, has the form of the following evolutionary equation
\begin{equation*}
\iun\hslash \frac{\partial \psi }{\partial t}=\widehat{H}\psi 
\end{equation*}
%
with the Hamiltonian, energy operator, given by
\begin{equation*}
\widehat{H}=\frac{{\widehat{\mathbf{p}}}^{2}}{2m}+U\left( \mathbf{r}\right) ,\qquad {\widehat{\mathbf{p}}}=%
\frac{\hslash }{\iun}\nabla =\frac{\hslash }{\iun}\frac{\partial }{\partial 
\mathbf{r}}\/,
\end{equation*}
%
and $U\left( \mathbf{r}\right)$ is the electron potential energy ($\hslash $ --  reduced Planck's constant).
As a result, the time-dependent Schr\"{o}dinger equation can be written as the partial differential equation
\begin{equation*}
\iun\hslash \frac{\partial \psi }{\partial t}=-\frac{\hslash ^{2}}{2m}\Delta
\psi +U\psi .
\end{equation*}

\medskip
\noindent
The (de Broglie) matter wave function $\psi= \psi(\mathbf{r},t)$ gives a complete mathematical description of the state of a quantum system.
Due to Born's {probabilistic interpretation}, the quantity $|\psi|^2$
presents the electron probability density with normalization%
\footnote{Schr\"{o}dinger originally proposed the quantity  $e|\psi|^2$ to represent the electron charge density (Figure~5).
But this interpretation would modify the classical Hamiltonian, resulting in a (nonlinear) integro-differential equation.
In the atom, classical concepts should be applied with caution, despite Bohr's correspondence principle.\/}  
\[
\int_{\mathbb{R}^3} |\psi|^2\,dv = 1 .
\]

\medskip
\noindent
If the potential is time independent $U = U(\mathbf r)$ we separate variables
\[
\psi(\mathbf r,t)
=
e^{-\iun Et/\hslash}\chi(\mathbf r)
\]
which leads to the {stationary Schr\"odinger equation}
\[
-\frac{\hslash^2}{2m}\Delta \chi + U\chi = E\chi ,
\]
or {eigenvalue problem} for the Hamiltonian
\[
\widehat{H}\chi(\mathbf{r}) = E\chi(\mathbf{r}) .
\]
This equation was first derived in the original publication \cite{SchrQMI} based on Hamilton's optical-mechanical analogy \cite{Mas:Ros, SchrQMII}.

\medskip
\noindent
According to Freeman Dyson \cite{Dyson2009}, 
one of the most profound `jokes of nature' is the square root of minus one that Erwin Schr\"{o}dinger had to put into his wave equation in 1926, when he invented wave mechanics.
And suddenly it became a new kind of wave equations instead of a heat conduction equation.

\medskip
\noindent
{\scshape{Remark:}} 
Utilization of the complex numbers was a hard decision for Schr{\"o}dinger to make. In the letter to Lorentz \cite[076\dag\;~p.~254]{Meyenn2011} he writes:
``...{\textit{The use of the complex number is unpleasant, indeed objectionable.
$\psi$
is therefore inherently a real function, so I should guess in equation
$(35)$ of my third paper $($\rm{\cite{SchrodingerCohrent}, SKS}$)$:}}
\begin{equation*}
\psi =\sum_{k}c_{k}u_{k}(x)e^{{\footnotesize{{\dfrac{2\pi \iun E_{k}t}{h}}}}}
\end{equation*}
\noindent
{\textit{
Instead of the imaginary exponent, I would like to neatly write a cosine and ask myself:
is it possible to define the imaginary part unambiguously without referring to the entire temporal course of the quantity, but only to the real quantity itself and its temporal and spatial differential quotients at the relevant point?}}
\footnote{{\textit{De facto\/,}} in order to obtain his nonspreading wave packets in closed form, Schr{\"o}dinger used the standard generating function for Hermite polynomials, which he found in \cite{Courant1924}.}

\medskip
\noindent
One must admit that the events that laid the foundations of modern wave mechanics developed extremely quickly a hundred years ago.
%
After discovering the relativistic version of his equation, presumably in late December of 1925 in Arosa, and failing to obtain the correct fine structure formula for a hydrogen atom \cite{SomAS}, Schr\"{o}\-dinger immediately switched to the non-relativistic case,
which since then bears his name
(see, for example, \cite{Barleyetal2021, Kragh1982, Kragh1984} for more details). 
As we understand nowadays, he invented a new class of partial differential equations, namely, Schr\"{o}\-dinger-type equations.
And the square root of minus one in the equation means that nature, for some mysterious reason, works not only with real numbers, but also with complex ones \cite{Dyson2009}. 
The further description of electron spin actually uses the concept of quaternions
in the form of Pauli matrices \cite{Jammer1989, Lan:Lif}. 
%

\medskip
\noindent
The stationary Schr\"{o}dinger equation for an electron in the central
field with the potential energy $U(r) $ is given by%
\begin{equation*}
\Delta \chi +\frac{2m}{\hbar ^{2}}\left[ E-U(r) \right] \chi =0
\label{s6}
\end{equation*}%
by the separation of variables \cite{Lan:Lif}. 
It turns out that this equation can be extended to multi-electron systems, which describes correctly everything we know about the behavior of atoms. 
It is also the basis of all of chemistry and most of quantum physics.
This discovery came as a complete surprise to Schr\"{o}dinger as well as to everybody else%
{\footnote{The connection between the Heisenberg and  Schr\"{o}dinger formulations of quantum mechanics and earlier work by
Lanczos based on an integral equation is discussed by van der Waerden \cite{vanderwarden}. 
According to \cite{Condon1962}, Born and Heisenberg consulted David Hilbert about their nascent matrix mechanics, and he pointed out
that their matrices could be a kind of by-product of the eigenvalues of the boundary-value problem
of a certain differential equation.
Hilbert later liked to mention that the wave equation could have been discovered at least six months earlier if they had followed his advice.
}}~%
.

\medskip
\noindent
{\scshape{Fundamental result:}}
But first, Schr\"{o}dinger found to his delight that his stationary wave equation has solutions corresponding to the quantized orbits in the Bohr model of the atom \cite{Bar:Ruf:Sus, Barleyetal2021, Dyson2009, Lan:Lif, SchrQMI}%
\footnote{
In a footnote on p.~3 of his first `quantization' article \cite{SchrQMI}, Schr\"{o}dinger writes: {``\it For guidance in the treatment of $($7$)$ $(${\rm radial equation, SKS}$)$ I owe thanks to Hermann Weyl.''}
Originally,  Schr{\"o}dinger found solutions by the Laplace contour integral method (see \cite{Barleyetal2021} for more details).
He also considered in detail the continuous spectrum, including the zero energy level, using the theory of analytic functions.
Completeness of solutions is also briefly mentioned. 
}%
.

\medskip
\noindent
In 1885 Swiss mathematician Johann Balmer discovered a remarkable numerical pattern
in the spectral lines of hydrogen (Figure~4).
The wavelengths satisfy the empirical relation

\[
\frac{1}{\lambda}
=
R
\left(
\frac{1}{2^{2}}-\frac{1}{n^{2}}
\right),
\qquad n=3,4,5,\dots \quad (R - {\rm{Rydberg\ constant\/}}).
\]

\begin{figure}[hbt!]
\centering
\includegraphics[width=1.0\textwidth]{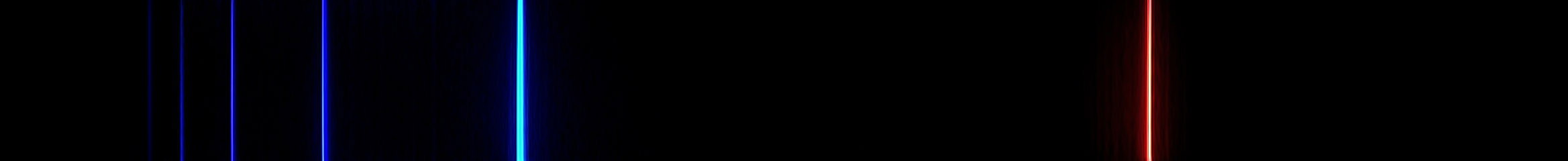}
\caption{
Balmer series -- the first four spectral lines of hydrogen (visible spectrum): \\
{\footnotesize
\textcolor{red}{$H_{\alpha}$} (\textcolor{red}{Red}): 
$\lambda = 656\,\text{nm}$ 
(transition $n = 3 \rightarrow n = 2$)\\[0.01cm]
\textcolor{LightSeaGreen}{$H_{\beta}$} (\textcolor{LightSeaGreen}{Blue--green}): 
$\lambda = 486\,\text{nm}$ 
(transition $n = 4 \rightarrow n = 2$)\\[0.01cm]
\textcolor{blue}{$H_{\gamma}$} (\textcolor{blue}{Blue}): 
$\lambda = 434\,\text{nm}$ 
(transition $n = 5 \rightarrow n = 2$)\\[0.01cm]
\textcolor{violet}{$H_{\delta}$} (\textcolor{violet}{Violet}): 
$\lambda = 410\,\text{nm}$ 
(transition $n = 6 \rightarrow n = 2$)}
\/.}
\label{FigureBalmer}
\end{figure}

\medskip
\noindent
Schr\"{o}dinger provided the theoretical foundation for these lines through his wave equation -- the progression from observing atomic patterns to deeply understanding the quantum behavior of electrons.

\medskip
\noindent
Developing the new wave mechanics, in the next few months, Schr\"{o}dinger himself, and/or his followers, had succeeded in solving long-standing problems in atomic physics. 
Among them were calculations of the Stark and Zeeman effects, describing behaviour of the hydrogen atom in electric and magnetic fields, respectively \cite{SchrQMI, SchrQMII, SchrodingerParabolic, SchrodingerCohrent, SchrodingerTimeDependent, SchrPhysRev1926}.
%
These papers, later combined in one volume and translated to English \cite{SchrodingerBook}, were originally written {\it{one by one\/}} at different times. 
The results of the further developments were largely unknown to the writer of the earlier ones.
There is no doubt that the efficient publication of these classical works remains a monumental `snapshot' in the history of science \cite{SchrQMI, SchrQMII, SchrodingerParabolic, SchrodingerCohrent, SchrodingerTimeDependent, SchrPhysRev1926} (see also {\url{https://schroedinger100.univie.ac.at/}}).
\medskip
\noindent
The international physics community reacted to Erwin Schr\"{o}dinger’s new wave mechanics with profound enthusiasm and a wave of relief, 
as his equation provided an intuitive, visual, and mathematically familiar alternative to the abstract matrix mechanics 
proposed earlier by Werner Heisenberg \cite{Bethe1977, Fock1926, MEHRAIII}.
Within a short time after the publication of Schrodinger's paper \cite{SchrQMI} wave mechanics was successfully applied to a great number of energy-eigenvalue problems \cite{EEKS, Flugge1999, Fock1926}.
Further development of quantum mechanics and its applications to molecular physics, to the solid state of matter, to liquids and gases, to statistical mechanics, as well as to nuclear physics, demonstrated the overwhelming generality of its methods and results
\cite{Jammer1989}.

\medskip
\noindent
Finishing the most productive year of his scientific career, Schr\"{o}dinger's review article \cite{SchrPhysRev1926} gives an account of the new form of quantum theory. That article was submitted to {\it Physical Review} on September~3, 1926, before visiting the United States later that year (Figure~5). 
At the end, addressing the controversy with the relativistic hydrogen atom, Schr\"{o}dinger concludes: ``{\textit{The deficiency must be intimately connected with Uhlenbeck--Goudsmit's theory of the spinning electron}}%
{\footnote{The concept of electron spin was introduced by G.~E.~Uhlenbeck and S.~Goudsmit in a
letter published in {\it{Die Naturwissenschaften\/}}; the issue of 20 November 1925 (see \cite{Jammer1989, Mehra1969} for more details\/).
And just a few days later, independently, according to
{\url{https://physik.uni-graz.at/en/news/happy-birthday-schroedingergleichung/}}
on November 23, 2025, Schr\"{o}dinger held at the Z\"{u}rich Polytechnic Institute his first colloquium talk 
on the work of Louis de Broglie.}}
~.{\textit{ But in what way the electron
spin has to be taken into account in the present theory is yet unknown".}}

\begin{figure}[hbt!]
\centering
\includegraphics[width=0.77\textwidth]{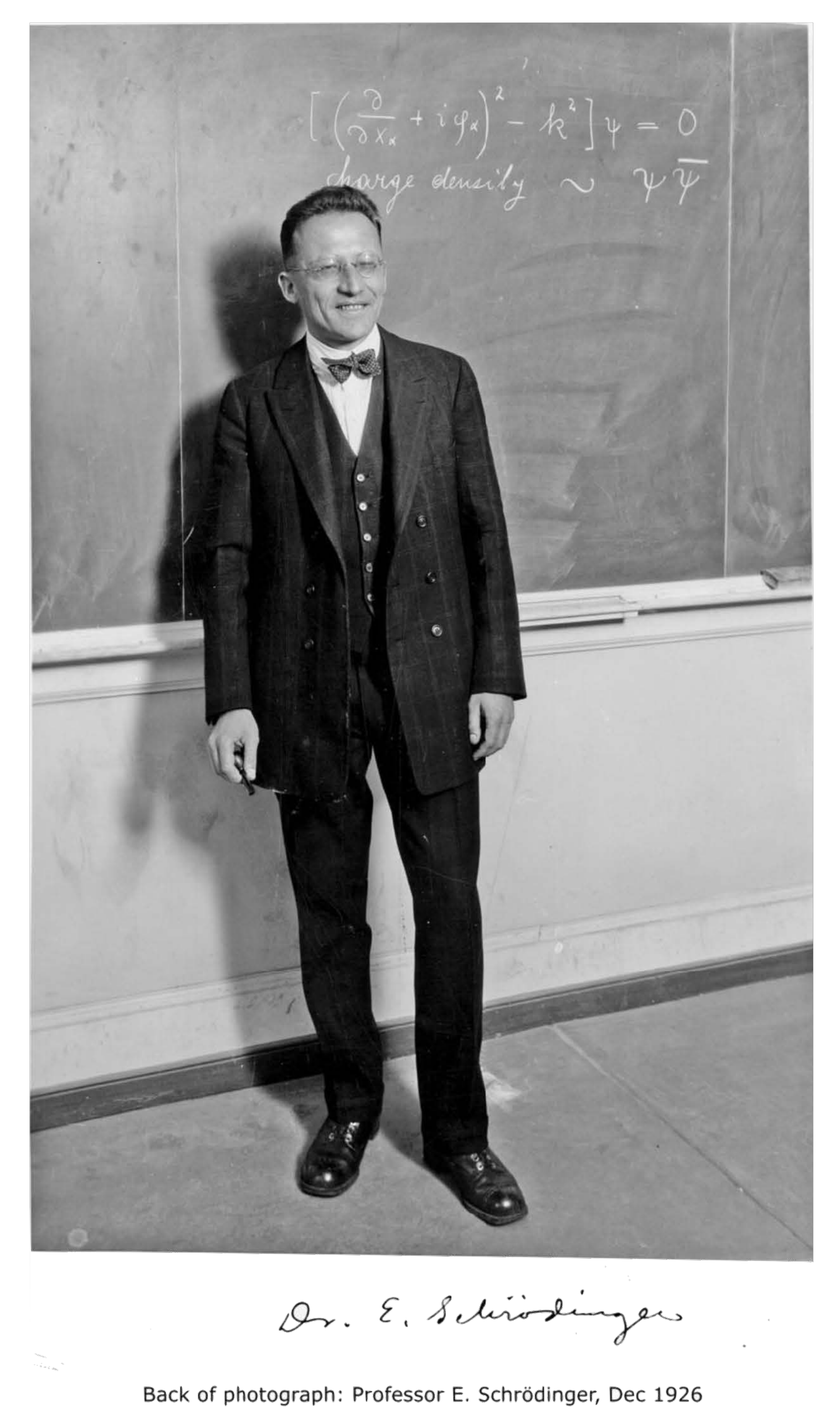}
\caption{
Erwin Schr\"{o}dinger accepted an invitation to lecture at the University of Wisconsin--Madison in early 1927, leaving in December 1926 to give talks in January and February 1927
{\url{https://search.library.wisc.edu/digital/AHDIU5YGAIRZOW8N}}
\/.}
\label{Figure5}
\end{figure}

\medskip 
\noindent
Less than two years later, Paul Adrien Maurice Dirac would derive a first-order wave equation for a four-component spinor field that describes relativistic spin-1/2 particles like electrons \cite{Dirac1928}, 
an early milestone on the long road to the Standard Model, 
the journey of upgrading from a single-particle quantum equation to a full Quantum Field Theory defined by local gauge symmetries \cite{Wilczek2004}.

%
\medskip
\noindent
{\scshape{Confession:}}
Thirty years later, in connection with Sommerfeld's fine-structure formula \cite{SomAS}, Erwin Schr\"{o}\-dinger testified, {\it inter alia},
in a letter dated 29 February 1956 \cite[pp.~113--114]{YouMan}:
{{\it{%
{\textquotedblleft ... you are naturally aware of the fact that Sommerfeld derivation of the fine-structure formula provides only fortuitously
the result demanded by the experiment.
One may notice then from this particular example that newer form of quantum theory $($i.~e., quantum mechanics\/$)$
is by no means such an inventible continuation of the older theory as is commonly supposed.
Admittedly the Schr\"{o}dinger theory, relativistically framed $($without spin\/$)$, gives a {\it{formal}} expression of the fine-structure formula
of Sommerfeld, but it is {\it{incorrect}} owing to the appearance of half-integers instead of integers.
My paper in which this is shown has ... never been published; it was withdrawn by me and replaced by non-relativistic treatment...
The computation $[$by the relativistic method\/$]$ is far too little known.
It shows in one respect how {\it{necessary}} Dirac's improvement was, and on the other hand it is wrong to assume that the older form of quantum theory is `broadly' in accordance with the newer form.%
\/\textquotedblright}}}} (See \cite{Bar:Ruf:Sus} for more details and also \cite[Appendix~D, p.~100]{Barleyetal2021} about the discovery of the relativistic Schr\"{o}dinger equation.)
\begin{figure}[hbt!]
\centering
\includegraphics[width=1.0\textwidth]{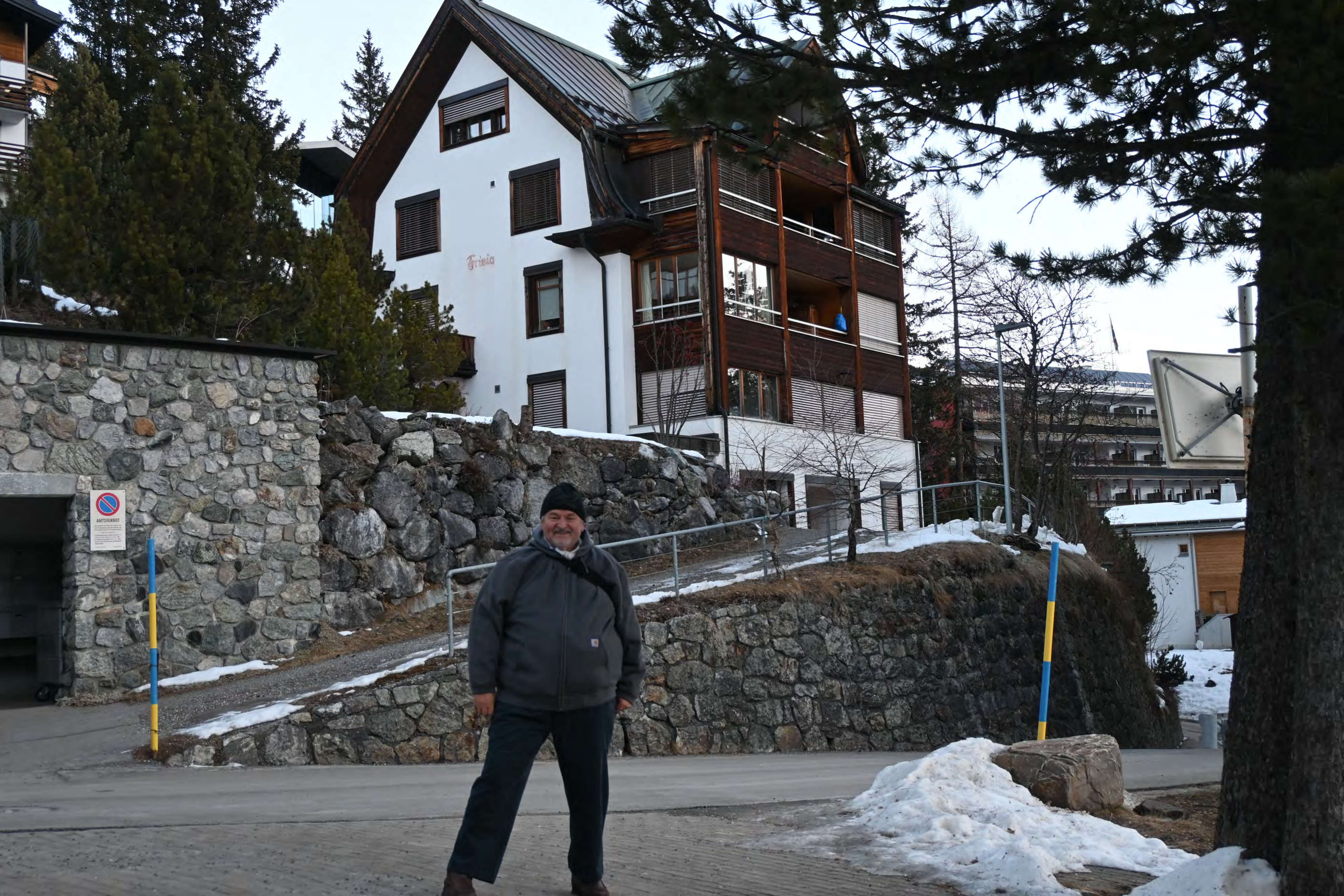}
\caption{
The author at Villa Frisia on December 29, 2025 
.}
\label{Figure6}
\end{figure}

\medskip
\noindent
{\scshape{Conclusion:}}
The year 1926 saw revolutionary change to the world of physics 
--- the Schr\"{o}\-dinger equation turned out to have an enormously wide range of applications, from the quantum theory of atoms and molecules to solid state physics, quantum crystals, superfluidity, and superconductivity.
In a historical perspective, the `golden years' \cite{Mehra2001} following the foundation of quantum mechanics have paved the way to quantum field theory and the physics of elementary particles.
Just in a few months, Erwin Schr\"{o}dinger turned the matter wave hypothesis of Louis de Broglie into a detailed, perfectly operating technique, which eventually eliminated many of `old' quantum physics postulates, that had been employed in the previous decades, 
showing they were unnecessary.

\medskip
\noindent
However, at that time, the `true nature' of quantum motion, including the probabilistic interpretation of the wave function, the uncertainty principle, the wave-particle duality, and the concept of spin, was not yet well-understood --- it came later as a result of vigorous debates in the quantum community which led to the creation of an intellectual structure of extraordinary beauty \cite{Jammer1989, MEHRAIII, Rechenberg}.%

\medskip
\noindent
Physicists concluded that subatomic particles do not possess simultaneous, exact values for position and momentum, nor do they strictly travel in continuous trajectories. Instead, quantum objects are governed by wave functions that define statistical probabilities.
Matter and radiation exhibit aspects of both localized particles and delocalized waves. 
The mathematical wave function does not represent a physical object and/or field, but rather the \textit{probability} of what an observer will find upon measurement.
Some debates about this so-called Copenhagen interpretation are not completed even 65 years after Schr\"{o}dinger! \cite{Jammer1989, Rechenberg}.

\medskip
\noindent
%
The fundamental paradigm of quantum mechanics, 
the revolutionary mathematical approach to describing quantum motion,  
lies in the recognition of the evolution 
of the state of a microscopic system as a dynamic process in an abstract (Hilbert) functional space (`quantum kinematics').
Treating quantum states as vectors in a complete inner-product space allows physicists to describe subatomic dynamics through the elegant mathematics of linear algebra and functional analysis.
%
The time-dependent Schr\"{o}dinger equation (`quantum dynamics') gave rise to a new class of partial differential equations
\cite{Dyson2009}.
Upon measurement, the wave function reduces to a single eigenstate corresponding to the measured value (`collapse of the wave function').
\medskip
\noindent
As a result, quantum mechanics has had a profound influence on many areas of mathematical physics and pure mathematics, including Sturm--Liouville theory and perturbations, orthogonal polynomials and special functions, operator algebras and spectral theory of operators on Hilbert spaces, axiomatic methods, probability and statistics, and the theory of group representations. These mathematical fields, in turn, provide the essential language and structure for modern physics. 
\medskip

%
\begin{center}
{\scshape{Happy Anniversary, Wave Mechanics!}}
\end{center}

\medskip
\noindent \textbf{Acknowledgments.\/}
The author is grateful to Kamal Barley, John Galvan, Sergey Kryuchkov, Nathan Lanfear, Andreas Ruffing, Eugene Stepanov, and Sergei Tabachnikov for their valuable comments and help.
\medskip
\noindent \textbf{Addendum.\/}
The author had a chance to visit Arosa, Switzerland, The ``Mecca of Quantum Physics'',
 on December 28--29, 2025 (see Figure~6).
\medskip 

{\center{\hrulefill}}

\bigskip
\noindent {\scshape{Extended Personal Dedication:\/}}
Max Planck, Ernest Rutherford, Albert Einstein, Niels Bohr, Arnold Sommerfeld, Louis de Broglie, Werner Heisenberg, Erwin Schr\"{o}dinger, and Paul Dirac revolutionized physics. In the early 20th century, they developed quantum theory, which changed our understanding of matter, light, and the subatomic world.
For our generations of physicists, they have become ``legends of quantum science'' on the scientific Olympus.
There are amazing personal connections, which
bridges decades of mathematical physics, the international dissemination of quantum mechanics, and a wonderful cross-generational academic legacy.
\smallskip

\noindent 
I would like to dedicate this essay to Professor Alladi Ramakrishnan and his son, Krishna Alladi. 
One can position Alladi Ramakrishnan (1923–2008), who was born at the same time as de Broglie's hypothesis about the wave properties of matter, 
within the second generation of quantum physicists.
He made pioneering contributions to cosmic radiation, cascade theory, matrices, quantum electrodynamics 
and acted as a grand host and bridge to the first generation. In 1960, Ramakrishnan hosted Niels Bohr at his family home, Ekamra Nivas, 
in Madras (Figure~7).
This famous ``Theoretical Physics Seminar'' directly inspired Ramakrishnan to found MATSCIENCE (The Institute of Mathematical Sciences) in India in 1962, institutionalizing quantum and mathematical research for future generations.
\vfill
\eject
\newpage 

\noindent 
His son, Professor Krishna Alladi, represents the natural progression of this lineage into modern mathematics.
He went on to become a world-renowned mathematician specializing in number theory (coauthor of Paul Erd\H{o}s and the founder of The Ramanujan Journal). 
Thanks to him, the legacy of rigorous mathematical structures that underlie modern physics continues to flourish in the Alladi family.

\smallskip    
\noindent 
This essay is warmly dedicated to the memory of Professor Alladi Ramakrishnan as well as to the 70th birthday of his son, Professor Krishna Alladi, celebrating the beautiful passing of the scientific torch across generations. 
This global legacy is mirrored profoundly by the intellectual heritage of the author. 
\smallskip

\noindent   
At the dawn of the 1960s, the great patriarch of quantum physics, Niels Bohr (1885--1962), made two historic journeys: one to Madras in 1960, where he inspired the Alladi family and the creation of MATSCIENCE, and another to Moscow/Dubna in 1961, where, among other things, he met with my future Ph.D. advisor, Professor Yakov A. Smorodinsky  (Figure~8), who was former Ph.D. student of Lev Landau (Nobel Prize in Physics, 1962):
{\url{http://theor.jinr.ru/~kuzemsky/yaasmor.html}}.
In honoring of these parallel origins, this dedication aims to highlight the deep scientific and human connections that unite the early pioneers of the quantum revolution with modern generation in mathematics and physics.
\smallskip

\noindent 
Niels Bohr's final visits to India (January 1960) and the USSR (May 1961) marked the twilight of his life and served as major bridges between Western quantum physics and the developing or communist academic worlds.
\bigskip

\begin{figure}[hbt!]
\centering
\includegraphics[width=1.0\textwidth]{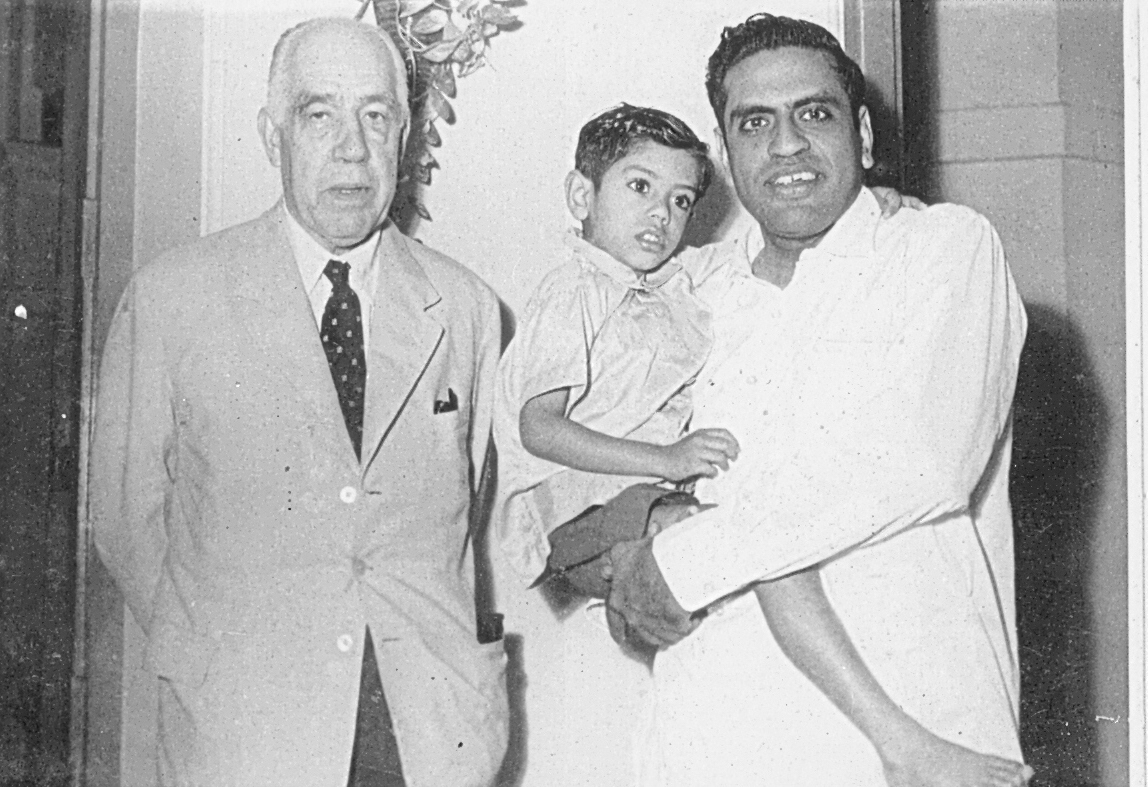}
\caption{
Krishna Alladi with his father Alladi Ramakrishnan 
and Nobel Laureate Niels Bohr at Ekamra Nivas, January 1960%
.}
\label{Figure7}
\end{figure}
\begin{figure}[hbt!]
\centering
\includegraphics[width=1.0\textwidth]{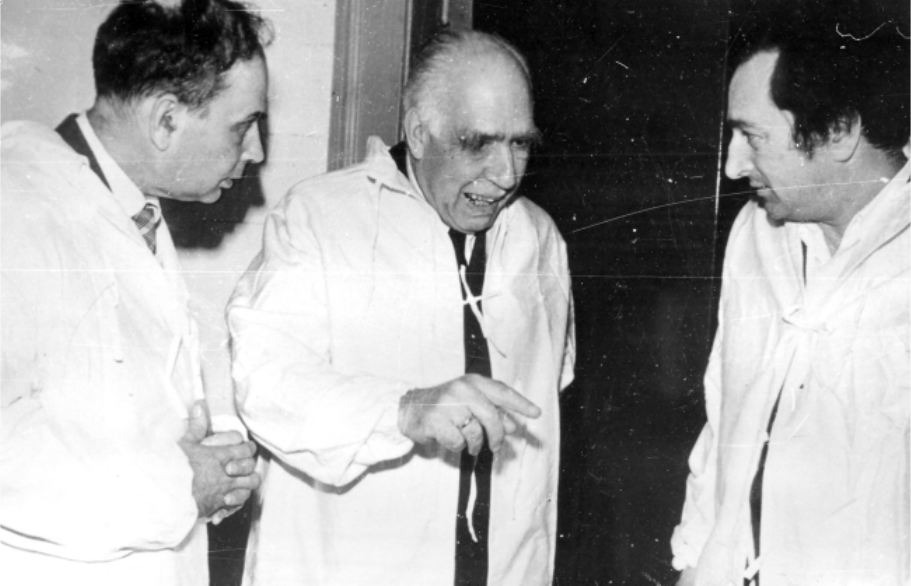}
\caption{
Nobel Laureates Ilya~M. Frank and Niels Bohr, along with 
Professor Yakov~A.~Smorodinsky, discussing physics of a new research reactor at the Joint Institute for Nuclear Research, Dubna, USSR, May 1961.
{\url{https://arkiv.dk/vis/5911192}};
{\url{https://www.jinr.ru/about-en/history-jinr/jinr-chronology/}}
}
\label{Figure8}
\end{figure}

\vfill
\eject
\newpage


%
\bigskip
{\url{https://orcid.org/0000-0001-8169-0987}}
\medskip
%
\end{document}